\documentclass[aps,prl,superscriptaddress,twocolumn,longbibliography]{revtex4-2}
\usepackage{graphicx}
\usepackage[hidelinks]{hyperref}
\usepackage{xcolor}
\usepackage{amsmath}
\usepackage{amsfonts}
\usepackage{amssymb}
\usepackage{varioref}
\usepackage{float}
\usepackage{dcolumn}   
\usepackage{subfigure}
\usepackage{bm}
\usepackage{multirow}
\usepackage[normalem]{ulem}

\usepackage{epstopdf}
\newcommand{\bew}{\begin{widetext}}
\newcommand{\ew}{\end{widetext}}

\newcommand{\beq}{\begin{equation}}
\newcommand{\eeq}{\end{equation}}
\newcommand{\beqn}{\begin{eqnarray}}
\newcommand{\eeqn}{\end{eqnarray}}

\newcommand{\dd}{{\rm d}}

\newcommand{\fig}{Fig.\ }

\let\oldepsilon\epsilon
\let\epsilon\varepsilon
\let\varepsilon\oldepsilon

\def\0v{\mathbf{0}}

\def\rv{\mathbf{r}}

\def\ev{\mathbf{e}}

\begin{document}

\title[]{Active Jamming at Criticality}

	\author{Shalabh K. Anand}
	\affiliation{Department of Bioengineering, Imperial College London, South Kensington Campus, London SW7 2AZ, U.K.}
	\affiliation{Department of Mathematics, Imperial College London, South Kensington Campus, London SW7 2AZ, U.K.}
	\author{Chiu Fan Lee}
	\email{c.lee@imperial.ac.uk}
	\affiliation{Department of Bioengineering, Imperial College London, South Kensington Campus, London SW7 2AZ, U.K.}
	\author{Thibault Bertrand}
	\email{t.bertrand@imperial.ac.uk}
	\affiliation{Department of Mathematics, Imperial College London, South Kensington Campus, London SW7 2AZ, U.K.}

\date{\today}

\begin{abstract} 
Jamming is ubiquitous in disordered systems, but the critical behavior of jammed solids subjected to active forces or thermal fluctuations remains elusive. In particular, while passive athermal jamming remains mean-field-like in two and three dimensions, diverse active matter systems exhibit anomalous scaling behavior in all physical dimensions. It is therefore natural to ask whether activity leads to anomalous scaling in jammed systems. Here, we use numerical and analytical methods to study systems of active, soft, frictionless spheres in two dimensions, and elucidate the universal scaling behavior that relates the excess coordination, active forces or temperature, and pressure close to the athermal jammed point. We show that active forces and thermal effects around the critical jammed state can again be captured by a mean-field picture, thus highlighting the distinct and crucial role of amorphous structure in active matter systems. 
\end{abstract}

\maketitle

Disordered systems of particles are ubiquitous in nature, from molecular glasses, colloidal suspensions and foams to biological tissues and granular materials. As such, models of jammed solids have naturally attracted intense attention from both theorists and experimentalists	 alike~\cite{Cates1998,Liu1998,Kadanoff1999,Liu2010,Ohern2002,Ohern2003,Silbert2005,Xu2005a,Majmudar2007,
Heussinger2009,Chaudhuri2010,Drocco2005,Ono2002,Olsson2007,Zhang2009,Schreck2011,Ikeda2013b,Bertrand2014,Ikeda2015,Degiuli2015,Clark2018,Thompson2019,Vanderwerf2020,Wang2021,VanHecke2010,Bi2015}. For an athermal system of soft frictionless spheres, it is by now well established that upon increasing its density, the system generically develops a yield stress under which the solid respond elastically~\cite{Liu2010,Ohern2002,Ohern2003,Silbert2005,Xu2005a,Majmudar2007,Heussinger2009,Chaudhuri2010,Drocco2005}. This jamming transition, known as ``point $J$'', is characterized by a critical packing fraction $\phi_c$, at which the jammed solids is marginally stable~\cite{Cates1998,Ohern2003,Liu2010} and displays an isostatic contact network~\cite{Ohern2003}.  

Intriguingly, the mechanical properties of such a system exhibit critical scaling in the vicinity of point $J$~\cite{Goodrich2016,Liarte2019,Sartor2021}, and the corresponding upper critical dimension, $d_u$, is believed to be 2~\cite{Liu2010,Charbonneau2011,Charbonneau2012,Goodrich2012,Charbonneau2015,Goodrich2016,Charbonneau2017} Importantly, the fact that $d_u =2$ has  enabled researchers to successfully use mean-field theory to elucidate certain critical behavior of jamming for physically relevant dimensions ($d = 2, 3$). 

Like jamming, active matter has received much attention over the past two decades and has proved to be a fertile ground for novel physics~\cite{Ramaswamy2010,Marchetti2013,Cates2015,Romanczuk2012,Solon2015,Bechinger2016,Fily2012,Redner2013,Vicsek1995,Bricard2013,Bertrand2022,Lee2013,Vladescu2014,Bertrand2018a,Bertrand2018b,Mahmud2009}. Besides being of an intrinsic interest to physicists, active matter physics offers  quantitative descriptions of complex biological processes such the dynamics of cellular tissues in health and diseases or embryogenesis~\cite{Sadati2013,Bi2015,Bi2016,Park2015,Garcia2015,Malinverno2017,Mongera2018,Atia2018,Yan2019,Mitchel2020,Petridou2021,Grosser2021}. 
In contrast to athermal jamming, mean-field theory seems to {\it always} break down in {\it all} physical dimensions  in diverse active matter systems at criticality and in the order phases (see  \cite{chenCriticalphenomenon2015,tonerSwarmingDirt2018,chenMovingReproducing2020,chenPackedSwarms2022,zinatiDensepolar2022,Jentsch2023} for some recent examples.) Therefore, it is natural to ask whether adding activity to static jammed packings would alter the mean-field picture that governs its critical behavior.

While jamming under activity has recently been explored, existing works tend to focus on the glassy regime away from the critical point~\cite{Henkes2011,Berthier2013,Flenner2016,Berthier2017,Berthier2019,Klongvessa2019,Henkes2020,Loewe2020,Yang2022,Keta2022}, such as aging dynamics in active glasses~\cite{Nandi2018,Janssen2019,Mandal2020b,Mandal2021,Janzen2022} and connections between dense active systems and sheared athermal passive jammed solids~\cite{Agoritsas2019a,Agoritsas2019b,Agoritsas2021,Mo2020,Morse2021}, mirroring the earlier effort to recast shear induced fluctuation into an “effective temperature” in passive jammed systems~ \cite{Berthier2002,Makse2002,Ono2002,Xu2005b}. 
This is surprising since the jammed point has long been viewed as an end point of glassy phase~\cite{Liu1998,Liu2010,Parisi2010}. While recent studies have clarified the important differences between jamming and glass transitions for passive systems~\cite{Krzakala2007,Parisi2010,Chaudhuri2010,Jacquin2011,Ikeda2013a,Charbonneau2017}, this has however not been done for active systems.

Interestingly, a recent study revealed that systems of active Brownian particles (ABP) at high density jam intermittently and that the lifetimes of these transiently jammed states lengthens with the persistence of the particles~\cite{Mandal2020a}. Since the lifetimes of these transient jammed states can be arbitrarily long, they are clearly experimentally relevant. However, the associated scaling behavior remains unexplored. In this Letter, we fill this void by elucidating the scaling behavior of these transiently jammed states under both activity and thermal fluctuations around the static critical jamming point.

{\it Scaling ansatz.}---Starting from the established fact that passive and athermal jamming is a critical phenomenon for which the critical point is exactly at isostaticity (i.e., the average coordination $z$ is $z_{\rm iso} = 2d$), analysis based on scaling ansatz have provided scaling relations among diverse physical observables \cite{Goodrich2016}. Here, we also start by formulating a scaling ansatz that incorporates the effects of active forces in the athermal jamming scenario. Since prior work has demonstrated that active forces have the generic effects of unjamming a system, we start by considering a generic system of soft frictionless spherical particles above jamming onset, i.e. with pressure $p>0$. In the absence of active forces, the following scaling relation between $\Delta z \equiv z-z_{\rm iso}$ and $p$ is well established in 2D systems~\cite{Goodrich2016}:
\beq
\label{eq:zp}
\Delta z \sim p^{1/1.88}\ .
\eeq
We now imagine that the particles can exert a persistent active force $f$ in a randomly chosen direction. As adding this force will tend to unjam the system, we anticipate that $\Delta z$ should decrease with $f$. At the same time, we expect the system to revert back to the same inherent structure when active forces are removed as long as $\Delta z$ remains positive. In other words, we expect the system to only explore locally its potential energy landscape and to remain in the basin of attraction of its initial static configuration. This expectation is justified by previous numerical simulations \cite{Mandal2020a} and our own simulation results. 

Around this critical region where $\Delta z$, $p$, and $f$ are all small, we expect that the system to be scale invariant with a single length scale that is controlled by one of the control parameter \cite{Ellenbroek2006}; we choose our control parameter to be $p$ in this case since it is the physical quantity that we actually control in most simulation studies. As such, we arrive at the following scaling ansatz:
\beq
\label{eq:scaling}
\Delta z = p^{1/1.88} S( f/p^\alpha)\ ,
\eeq
where $S$ is a {\it universal} scaling function such that $\lim_{x \rightarrow 0} S(x)$ is a positive constant so that we recover the static scaling relation between $\Delta z$ and $p$ (\ref{eq:zp}).

Again, since increasing $f$ has the generic effect of unjamming the system, we anticipate that the scaling function $S(x)$ is monotonically decreasing with respect to $x$, and eventually becomes 0 at $x_c$. In other words, there exists a critical force $f_{c}$ such that when 
\beq
\label{eq:fc}
f_{c} = x_c p^\alpha\ ,
\eeq
$\Delta z$ is exact zero in the system. We now test this scaling ansatz via means of numerical simulations.

\begin{figure}[t!]
\centering
\includegraphics[width=0.45\textwidth]{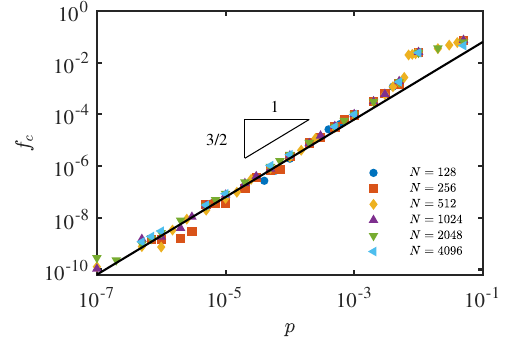}
\caption{{\it Scaling of the critical active force ($f_c$) vs.~pressure ($p$).} The critical force $f_c$ is defined as the active force  at which the contact network of an initially over-jammed system (with pressure $p$) becomes isostatic. The scaling behavior found in our ABP systems for various system sizes $N \in [128,4096]$ is consistent with a power-law scaling with exponent $\alpha = 3/2$ (solid black line).}
\label{fig:crit_force}
\end{figure}

\textit{Model and simulations.}---We study the dynamics of two-dimensional ($d=2$) jammed packings of $N$ spherical particles interacting via purely repulsive harmonic forces when subjected to active forces and thermal fluctuations. We start by creating static jammed packings at fixed pressures $p$ providing a way to control the packings distance to the athermal jamming transition point. To do so, we start with random configurations of discs at high volume fraction ($\phi=0.95$) in a square box of size $L$ with periodic boundary conditions. We then follow the FIRE (Fast Inertial Relaxation Engine) energy minimization procedure until the maximum unbalanced force on any particle is less than $f_t = 10^{-14}$. Following each energy minimization step, we either increase (if the pressure is below target) or decrease (if the pressure if above target) the diameter of the particles proceeding to a bisection until we obtain a mechanically stable configuration within $1\%$ of the target pressure $p$. To avoid crystallization in the system, we work with $50:50$ binary mixtures of particles with diameter ratio $1:1.4$ as is typically done in studies of disordered solids \cite{Ohern2003,Xu2005a}. 

These mechanically stable configurations form the initial conditions of our numerical simulations. Here, the time evolution of the positions of the particles is generically governed by an overdamped Langevin equation of the form
\begin{subequations}
\begin{align}
	\dot{\rv}_i &= \frac{1}{\gamma}\left[ -\sum_{j=1}^{N}{\bf \nabla}_{i}U({r}_{ij})+f{\bf{\hat e}}_{i} \right] + \sqrt{2D} \bm{\eta}_i \label{eq:langevineq_1}\\
	\dot{\theta}_{i} &= \sqrt{2D_{r}}\xi_{i} \label{eq:langevineq_2}
\end{align}
\label{eq:langevineq}%
\end{subequations}
where $r_{ij} = |\rv_i - \rv_j|$ is the distance between particles $i$ and $j$ and $\hat{\ev}_{i}=(\sin \theta_{i}, \cos \theta_{i})$ denotes the direction of self-propulsion, $f$ is the strength of active force and $\gamma$ is a friction coefficient. Both $\bm{\eta}_i$ and $\xi_i$ are zero-mean, unit variance Gaussian random variables such that 
\begin{subequations}
\begin{align}
	\langle \eta_{i,\alpha} (t) \rangle = 0~,& \quad \langle \eta_{i,\alpha}(t) \eta_{j,\beta}(t') \rangle = \delta_{i,j} \delta_{\alpha,\beta} \delta(t-t')\label{eq:noise_1}\\
	\langle \xi_{i} (t) \rangle = 0~,& \quad \langle \xi_{i}(t) \xi_{j}(t') \rangle = \delta_{i,j} \delta(t-t')	\label{eq:noise_2}
\end{align}
\label{eq:noise}%
\end{subequations}
with $(i,j) \in [1,N]$ and $(\alpha,\beta) \in \{x,y\}$. The direction of the self-propulsion is thus governed by rotational diffusion with coefficient $D_{r}$. Finally, $U(r_{ij})$ denotes the pairwise particle interactions which we model through a purely repulsive linear spring potential such that 
\begin{equation}
	U(r_{ij}) = \frac{\epsilon}{2} \left(1-\frac{r_{ij}}{\sigma_{ij}}\right)^{2} \Theta(\sigma_{ij}-r_{ij}), \label{eq:potential}	
\end{equation}
where $\sigma_{ij}=(\sigma_{i}+\sigma_{j})/2$ is the average diameter of particles $i$ and $j$ and $\Theta(x)$ is the Heaviside function. In what follows, we compare two main scenarios: (i) subjecting static jammed packings to persistent active forces via an ABP model, corresponding to solving Eq.\,(\ref{eq:langevineq}) with $D=0$ and $f>0$ and (ii) subjecting static jammed packings to thermal fluctuations via Brownian dynamics, corresponding to solving Eq.\,(\ref{eq:langevineq}) with $D>0$ and $f=0$. Finally, we also perform simulations for active Ornstein Uhlenbeck particles (AOUP) particles (see \footnote{See Supplemental Material at []} for details).

We nondimensionalized the equations of motion using the smaller particle diameter $\sigma$ and potential energy prefactor $\epsilon$ as basic units of length and energy. In ABP simulations, the typical timescale is given by the persistence time $\tau_p = D_r^{-1}$, while in Brownian dynamics simulations, a more natural choice for the timescale is given by $\tau = \sigma^2/D$. In all simulations, we set $\sigma=1$, $\epsilon=1$ and $\gamma=1$. Further, the rotational diffusion coefficient $D_{r}$ is taken to be $10^{-2}$ unless stated otherwise. To discard any potential finite size effect in our results, we varied the number of particles in the system from $N=128$ to $N=4096$. The target pressures of the initial configurations were taken in the range $[10^{-7}, 10^{-1}]$. Finally, results are averages over $100$ independent realizations.

To estimate the scaling exponent $\alpha$ using Eq.~(\ref{eq:fc}), we start with static packings above jamming onset with different pressure values $p$ in the absence of active forces; then increasing the active force strength, we find the critical value of the active force for which the time-averaged excess coordination $\Delta z_i$ becomes zero. Fig.\,\ref{fig:crit_force} shows that 
\begin{equation}
\alpha =1.5 \pm 0.1 \ .
\label{eq:alpha}
\end{equation}
We now present a heuristic argument that may justify the value of $\alpha$ obtained from our simulation.

\begin{figure}[t!]
	\centering
	\includegraphics[width=0.45\textwidth]{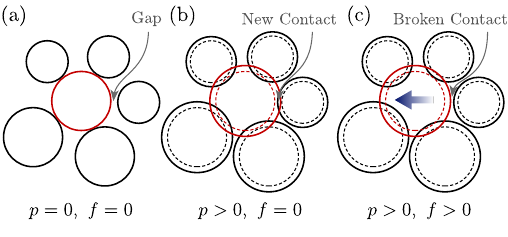}
	\caption{{\it Schematics illustrating our heuristic argument for value of $\alpha$.} (a) Particle configuration at critical jamming under no active force. The gap distribution between non-touching neighbors is expected to follow the scaling law in (\ref{eq:h}). (b) Upon further compression (or equivalently, upon particles' expansion), the pressure $p$ becomes positive and new contacts are formed as gaps close because the particles move closer to each other. The new particles' profiles are shown in solid lines whereas the pre-compressed profiles are in dashed lines. (c) When active forces are switched on $(f>0)$, contacts can be broken again as the active force (blue arrow) of the red particle can counteract the steric interactions with its neighbors.}
	\label{fig:cartoon}
\end{figure}

\begin{figure}
	\centering
	\includegraphics[width=0.95\columnwidth]{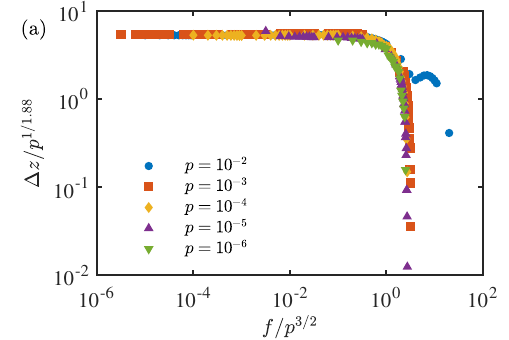} \\
	\includegraphics[width=0.95\columnwidth]{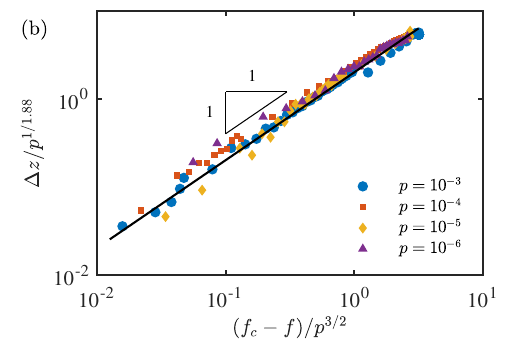}
	\caption{{\it Universal scaling function $S(x)$.} (a) Collapsed data for excess number of contacts per particle ($\Delta z$) as a function of the active force $f$ at various pressure values for $N=2048$. A good collapse is obtained via a rescaling of $\Delta z$ by $p^{\beta}$ (with $\beta =1.88$ for $d=2$, see \cite{Goodrich2016}) and $f$ by $p^\alpha$ with $\alpha = 3/2$. (b) Evolution of the excess coordination number $\Delta z$ near the critical active force for various values of pressure for $N=2048$. The black solid line shows a linear scaling. }
	\label{fig:scaling_function}
\end{figure}
	
{\it Heuristic argument for estimating $\alpha$.}---Under active forcing, the unjamming transition is generically controlled by the excess coordination $\Delta z$; it thus makes sense to expect the value of the critical active force to be dictated by the statistics of interparticle contacts in the overjammed static configurations. The scaling behavior of the distribution of interparticle gaps $h$ has been studied both analytically and numerically~\cite{Charbonneau2012,Wyart2012,Lerner2013,Charbonneau2017,Parisi2020,Babu2022}. Specifically, the probability distribution of $h$ in the vicinity of the athermal jamming point is given by 
\beq
\label{eq:h}
P(h) \sim h^{-\gamma}
\ .
\eeq
In our soft sphere simulations, we obtain $\gamma = 0.42 \pm 0.04$ \cite{Note1} which is close to the value of $\gamma_\infty = 0.41269...$ predicted in the mean-field theory of the glass transition in hard spheres in the limit of infinite dimensions \cite{Parisi2010,Charbonneau2012,Parisi2020}. In our model of soft spheres, the gap distribution shows a weak dependence on the pressure, which we neglect here as it does not affect our argument (at least within the precision of our numerical estimate (\ref{eq:alpha})).

We  now assume that the gap distribution remains the same at small enough pressures $p$ and active forces $f$. As the pressure $p$ increases in the absence of active forces, some of these gaps close and new contacts form (see cartoon in fig.~\ref{fig:cartoon}). Such an increase in new contacts is given by
\beq
\Delta z \sim \int_0^{\delta} h^{-\gamma} \dd h \sim \delta^{0.58}
\eeq
where $\delta$ is the average overlap. Note that since $\delta$ is known to scale like $p$ in the overjammed regime \cite{Parisi2020}, the numerical value is already in good agreement with the known scaling law between $\Delta z$ and $p$ (\ref{eq:zp}).

We now consider that each particle is subject to an active force $f$. To eliminate all the newly formed contact and thus bring the system back to the isostaticity, the average force needed is
\beq 
f_{c} \sim \int_0^\delta hP(h) \dd h \sim \delta^{1.58} \sim p^{1.58}
\ ,
\eeq
where the upper-bound $\delta$ again corresponds to the limit at which $N\Delta z$ contacts are eliminated. This heuristic argument thus support the value of $\alpha$ (\ref{eq:alpha}) obtained from our simulation.

{\it Universal scaling function $S$.}--- This newly found scaling allows us to elucidate the form of the universal scaling function $S$ numerically. As presented in Fig.\,\ref{fig:scaling_function}(a), $S(x)$ is monotonically decreasing with $f$ for all pressures. We obtain $S(x)$ by collapsing the surface plot displayed in \cite{Note1}. Provided the scaling behaviors given in Eqs.\,(\ref{eq:zp}) and (\ref{eq:fc}), we attempt to collapse our data. In particular, as the excess coordination scales as $p \sim \Delta z^{1.88}$, we can use this information to scale the curves in the low activity regime. Further, using the scaling of the critical active force with pressure, we obtain a good collapse of $\Delta z$ as a function of $f$ for all pressure values as can be seen on Fig.~\ref{fig:scaling_function}(a). This collapse of our data supports the existence of scaling relations and existence of universal behavior in the vicinity of the jamming point, i.e. at low enough pressures.

{\it Left continuity of $S$ as it vanishes.}---Before going any further, it is important to discuss one of the conditions under which measuring the critical active force $f_c$ is meaningful. In the use of the scaling relation between $f_{c}$ and $p$ in Eq.~(\ref{eq:fc}) to obtain the scaling exponent $\alpha$ above, we have focused on the vanishing point $S$ (or, equivalently, the point where $\Delta z=0$). However, it is clear that the system can no longer be rigid when $\Delta z$ becomes negative (i.e. below isostaticity) and so the scaling ansatz cannot be correct in this negative $\Delta z$ regime. In other words, it is unclear whether the location when $S$ vanishes is continuous at all.  Fortunately, left continuity of $S$---which corresponds to $\dd S(x)/\dd x |_{x=x_{c,-}}$ being well-defined---is all that is required for the above argument to work. In particular, this condition implies the following scaling relation for $f$ close to $f_c$
\beq
\frac{\Delta z}{p^{1/1.88}} \sim A \times \left(\frac{f_{c}-f}{p^\alpha}\right)
\ ,
\eeq
where $A \equiv -\dd S(x)/\dd x |_{x=x_{c,-}}$ is a positive constant. We show in \fig \ref{fig:scaling_function}(b) that the above scaling relation is indeed satisfied; this, in turn, implies the left-continuity of $S$ at the vanishing point, justifying our procedure  to determine $\alpha$ (see Fig.\,\ref{fig:crit_force}).

{\it Scaling behavior and persistence.}---Studies of athermal jammed packings generically characterize jammed configuration as having a positive excess coordination. When analyzing systems subjected to active forcing, we chose to extend this criterion and define jammed systems as systems whose {\it steady-state time-averaged} excess coordination is positive. However, nothing prevents the systems coordination from falling below the isostatic limit at least transiently. As a consequence, our systems may transiently unjam before rejamming. To analyze this, we studied (i) distributions of Debye-Waller factors, (ii) mean square displacements and (iii) intermediate scattering functions in our packings under active forces. We note that strikingly both quantities display signatures of the critical transition at $f = f_c$ and show that structural rearrangements are rare even on timescales which are long compared to the active force persistence time $\tau_p = D_r^{-1}$ \cite{Note1,Keta2022}.

\begin{figure}
\centering
\includegraphics[width=0.95\columnwidth]{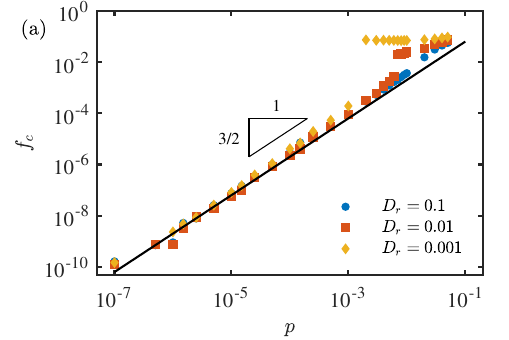} \\
\includegraphics[width=0.95\columnwidth]{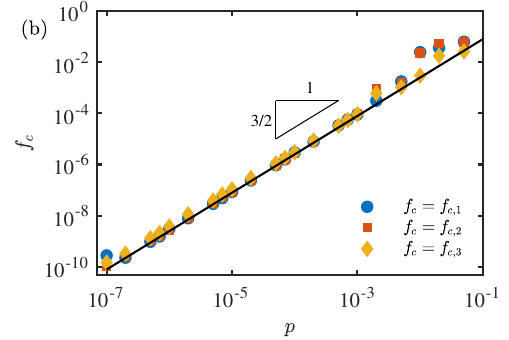}
\caption{{\it Effect of persistence and thermal fluctuations.} (a) Critical active force as a function of pressure for various values of the rotational diffusion constant $D_{r}$ in simulations of $N=512$ active Brownian particles (ABP). The solid black line shows a power law scaling with exponent $\alpha = 3/2$. For lower values of $D_r$, the self-propulsive forces become more persistent and the regime of validity of our power-law scaling is limited to lower values of the pressure. (b) Comparison of critical forcing for three perturbation protocols for systems of size $N=2048$: (i) persistent active forces with constant magnitude but diffusive direction in active Brownian particles simulations (ABP), (ii) thermal fluctuations in Brownian dynamics simulations (BD) and (iii) persistent active forces with fluctuating magnitude and direction in active Ornstein-Uhlenbeck particles simulations (AOUP). The critical forcing follows in all three cases the same scaling behavior as a function of $p$ close to the passive athermal jamming point ($p \to 0$). The solid black line shows a power law scaling with exponent $\alpha = 3/2$. }
\label{fig:persist_thermal}
\end{figure}

Having focussed on the long persistence time regime so far, we now consider how robust the scaling regime is as the persistence time of the active forcing decreases. Fig.\,\ref{fig:persist_thermal}(a) shows that the scaling behavior indeed persists, under the same measurement protocol, even as persistence length goes down by two orders of magnitude. Interestingly, we observe that as the persistence length of the particles decreases, the regime of validity of the critical force scaling shifts towards larger pressure values (as evidenced by the shoulders developing at the high $p$ limit), while scaling \eqref{eq:fc} remains intact in the vicinity of the jamming point.

{\it Active forcing vs.~thermal fluctuations.}---As we have seen that the scaling regime remains robust even for very small persistence lengths, it is thus natural to extend our analysis to the situation where jammed packings are subjected to thermal fluctuations (corresponding to the zero persistence time limit) and ask the question of whether the {\it same} scaling can also be measured robustly in this case. Although the jammed system will inevitably melt under thermal fluctuations in the long time limit, we will test the robustness of the scaling regime in the intermediate time regime (as defined within the same framework of our standard simulation and measurement protocols used so far).
 
To study thermal fluctuations, we switch off the activity and follow dynamically as above the network of contacts in Brownian dynamics simulations. We identify the critical temperature $T_c$ at which the time-averaged excess coordination $\Delta z$ vanishes. To be able to compare directly the effects of thermal fluctuations to that of active forces, we define here our critical forcing as $f_{c,2} = \sqrt{2T_c}$. Fig.\,\ref{fig:persist_thermal}(b) shows the critical thermal forcing $f_{c,2}$ as a function of the pressure $p$ in systems with $N=2048$ particles (see \cite{Note1} for a system size analysis). Strikingly, we show that the critical forcing for both active Brownian particles and their passive counterpart show the {\it same power-law scaling with pressure}. Finally, we also verify that we obtain the same scaling for AOUPs for which we estimate the critical force as $f_{c,3} = \sqrt{2D_{a}/\tau_a}$, where $\tau_{a}$ is the persistence time of the active force magnitude (for details see \cite{Note1}). We thus conclude to the existence of dynamically jammed states even at finite forcing whose scaling behavior is independent of the driving mechanism in the vicinity of point $J$.

{\it Summary \& outlook.}---In summary, we have studied the effects of active forces and thermal fluctuations on long-lived jammed states of soft frictionless particles in two dimensions close to the critical athermal jamming point. Using numerical simulations supplemented by a heuristic analytical argument, we elucidated the universal scaling relations between the excess coordination, the active force (or temperature), and pressure. In particular, the heuristic, mean-field based argument that we used to support the scaling exponent suggests that the mean-field picture applicable to athermal jamming continues to hold. This is in surprising contrast to diverse active matter systems in which anomalous scaling behavior is the norm. An interesting future direction would be to seek the existence of diverging length or time scales.

\begin{acknowledgments}
This work was supported by Leverhulme Trust (Grant: RPG-2019-374). Authors acknowledge the support of Research Computing Services at Imperial College London.
\end{acknowledgments}

%

\end{document}


\title{Supplemental Material for ``Active Jamming at Criticality"}

	\author{Shalabh K. Anand}
	\affiliation{Department of Bioengineering, Imperial College London, South Kensington Campus, London SW7 2AZ, U.K.}
	\affiliation{Department of Mathematics, Imperial College London, South Kensington Campus, London SW7 2AZ, U.K.}
	\author{Chiu Fan Lee}
	\email{c.lee@imperial.ac.uk}
	\affiliation{Department of Bioengineering, Imperial College London, South Kensington Campus, London SW7 2AZ, U.K.}
	\author{Thibault Bertrand}
	\email{t.bertrand@imperial.ac.uk}
	\affiliation{Department of Mathematics, Imperial College London, South Kensington Campus, London SW7 2AZ, U.K.}

\maketitle


\renewcommand{\theequation}{S\arabic{equation}}
\setcounter{equation}{0}

\renewcommand{\thefigure}{S\arabic{figure}}
\setcounter{figure}{0}

\section{AOUP Model}

To bridge the gap between our active Brownian particles simulations and our passive Brownian dynamics, we performed simulations for dense collections of active Ornstein-Uhlenbeck particles (AOUP). In this model, the position of the particles are governed by
\begin{equation}
\dot{\rv}_i = \frac{1}{\gamma} \left[ -\sum_{j=1}^{N}{\bf \nabla}_{i}U({r}_{ij})+\bf{f}_{i} \right]~.\label{eq:aoueq_1}
\end{equation}
where $\bf{f}_i$ denotes the active forces. While ABPs are subject to active forces with constant amplitudes and diffusing directions, AOUPs are subject to varying amplitude active forces which are governed by 
\begin{equation}
\tau_{a} \dot{\bf{f}}_{i} = -{\bf{f}}_i + \sqrt{2D_{a}} {\bf \psi}_i. \label{eq:aoueq_2}
\end{equation}
Here, $\tau_{a}$ is the persistence time of the active forces, $D_{a}$ a diffusion constant and ${\bf \psi }_{i}$ is a zero mean, delta correlated Gaussian white noise vector. We fix the diffusion coefficient $D_{a}$ and vary $\tau_{a}$ to change the strength of active force, defined in the AOUP model as $f = \sqrt{2D_{a}/\tau_a}$.

\section{Scaling relation between excess coordination and pressure}
\begin{figure}[ht!]
	\centering
	\includegraphics[width=0.48\columnwidth]{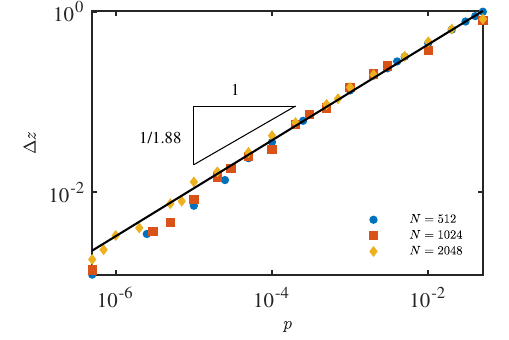}
	\caption{Excess coordination $\Delta z$ as a function of the pressure $p$. The solid line represents a power-law behavior with exponent $1/1.88$.}
	\label{fig:scaling_atherm}
\end{figure}
Recently, Goodrich et al. \cite{Goodrich2016} have argued that the excess coordination $\Delta z = z - z_{\rm iso}$ in static packings above the jamming threshold follows the following scaling relation with pressure 
\begin{equation}
\Delta z \sim p^{1/1.88}
\label{seq:scalingdz}
\end{equation}
Here, we confirm this scaling numerically by analyzing the static packings used as initial configurations for our numerical simulations. As presented in Fig. \ref{fig:scaling_atherm}, $\Delta z$ follows scaling relation (\ref{seq:scalingdz}).

\section{Gap distribution for static packings}

As argued in the main text, we expect the value of the critical active force to be dictated by the statistics of interparticle contacts in our overjammed static configurations. Those are in turn controlled by the distribution of gaps in static configurations at the jamming onset. As we overcompress our packings passed the jamming threshold, we expect gaps to close and new contacts to be formed. Conversely, as we increase the active force, we expect the particle coordination to decrease. 

\begin{figure}[h!]
	\centering
	\includegraphics[width=0.48\columnwidth]{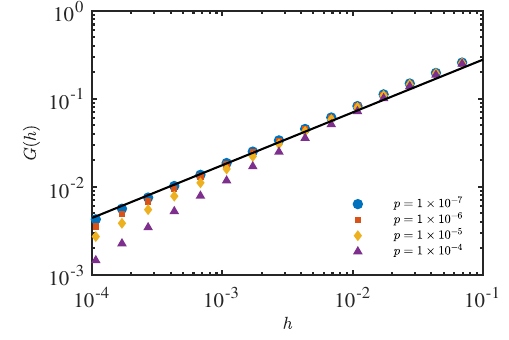}
	\caption{Cumulative distribution of gaps between particles in static packings at various pressures $p \in [10^{-7},10^{-4}]$ for system size $N=2048$.}
	\label{fig:gap_distribution}
\end{figure}

To show this, we compute the gap distribution for static packings of particles ($N=2048$); we expect the gap distribution to follow a power-law behavior 
\begin{equation}
P(h) \sim h^{-\gamma}.
\end{equation}
Figure\,\ref{fig:gap_distribution} shows that the cumulative distribution of gaps $G(h) = \int_{0}^{h}P(h')dh'$ does indeed display a power-law behavior
\begin{equation}
G(h) \sim h^{1-\gamma}
\end{equation}
with scaling exponent $\gamma = 0.42 \pm 0.04$ measured for $p=10^{-7}$. We note that the gap distribution is weakly dependent on the pressure $p$ itself.

\section{Excess contacts as a function of active force}

\begin{figure}[h!]
	\centering
	\includegraphics[width=0.4\columnwidth]{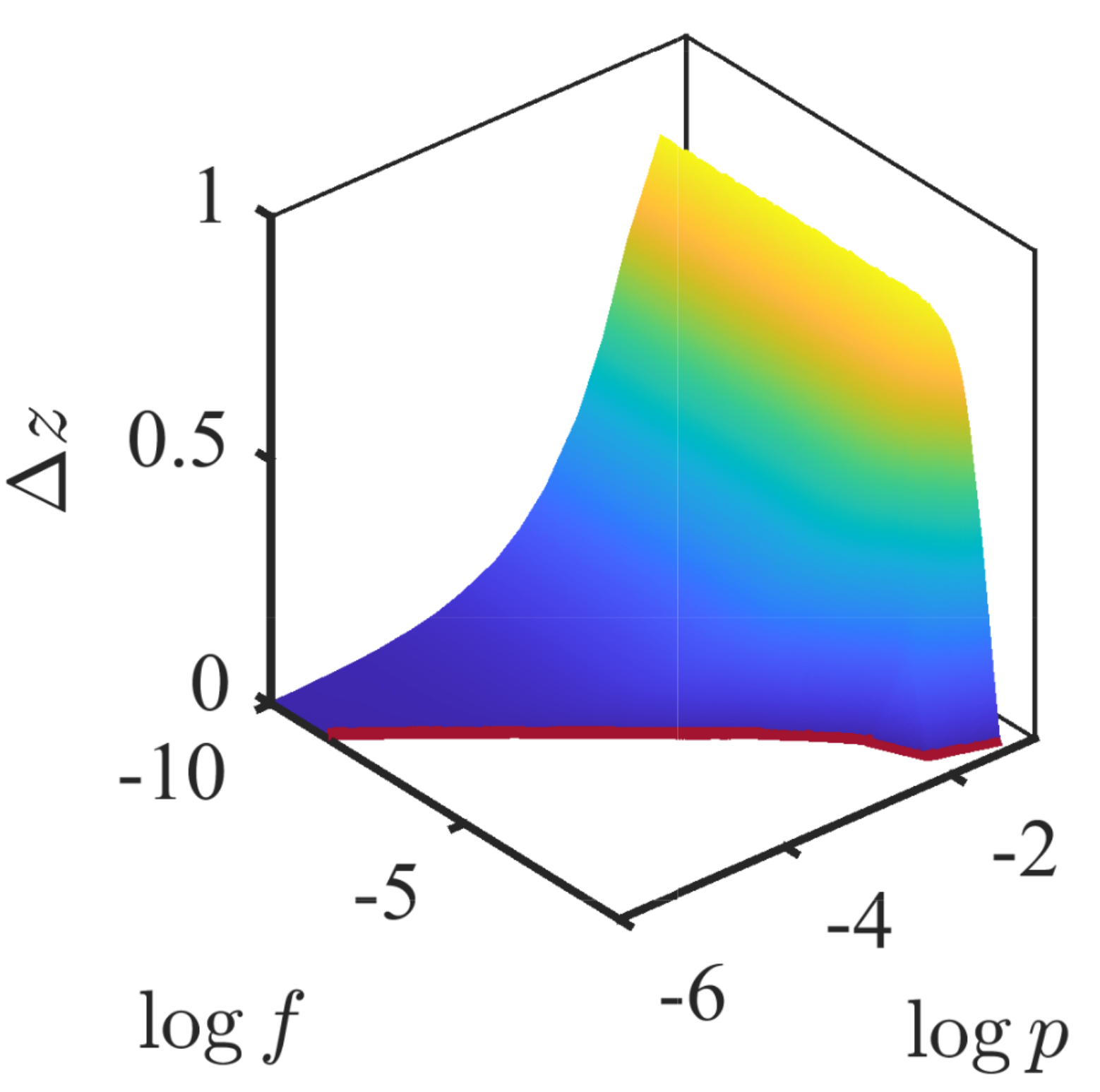}
	\caption{Excess coordination $\Delta z$ as a function of active force $f$ and pressure $p$. The red line represents the strength of active force at which $\Delta z$ becomes $0$. The colormap indicates the magnitude of $\Delta z$.}
	\label{fig:surface}
\end{figure}

We estimate $\Delta z$ for a series of active force strengths $f$ for a given pressure $p$. We represent the surface formed by the excess coordination for all active force/pressure pairs in Fig.\,\ref{fig:surface}. We observe a monotonic decrease in excess coordination with active forces. The excess coordination eventually becomes zero (displayed by red line in Fig.\,\ref{fig:surface}).

\section{System size effect on critical temperature}
\begin{figure}[h!]
	\centering
	\includegraphics[width=0.48\columnwidth]{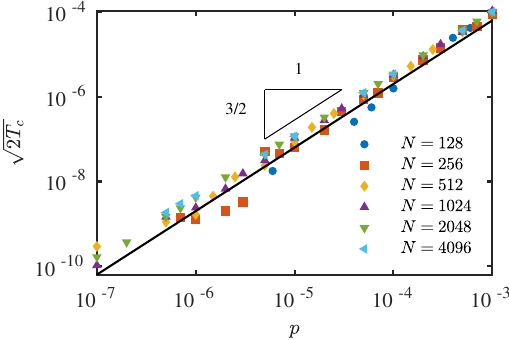}
	\caption{System-size effect on the critical temperature $T_c$ as a function of the pressure $p$.}
	\label{fig:thermal}
\end{figure}

In this section, we confirm the scaling of the critical temperature $T_c$ with pressure $p$ presented in Fig.\,3 for a variety of system sizes from $N=128$ to $4096$. We show here that the critical temperature follows the same scaling with the pressure $p$ for all system sizes tested (see Fig.\,\ref{fig:thermal}), namely 
\begin{equation}
f_{c,2} \equiv \sqrt{2T_c} \sim p^{3/2}
\end{equation}

\section{Dynamical signatures of unjamming}

\begin{figure}[t!]
	\centering
	\includegraphics[width=\columnwidth]{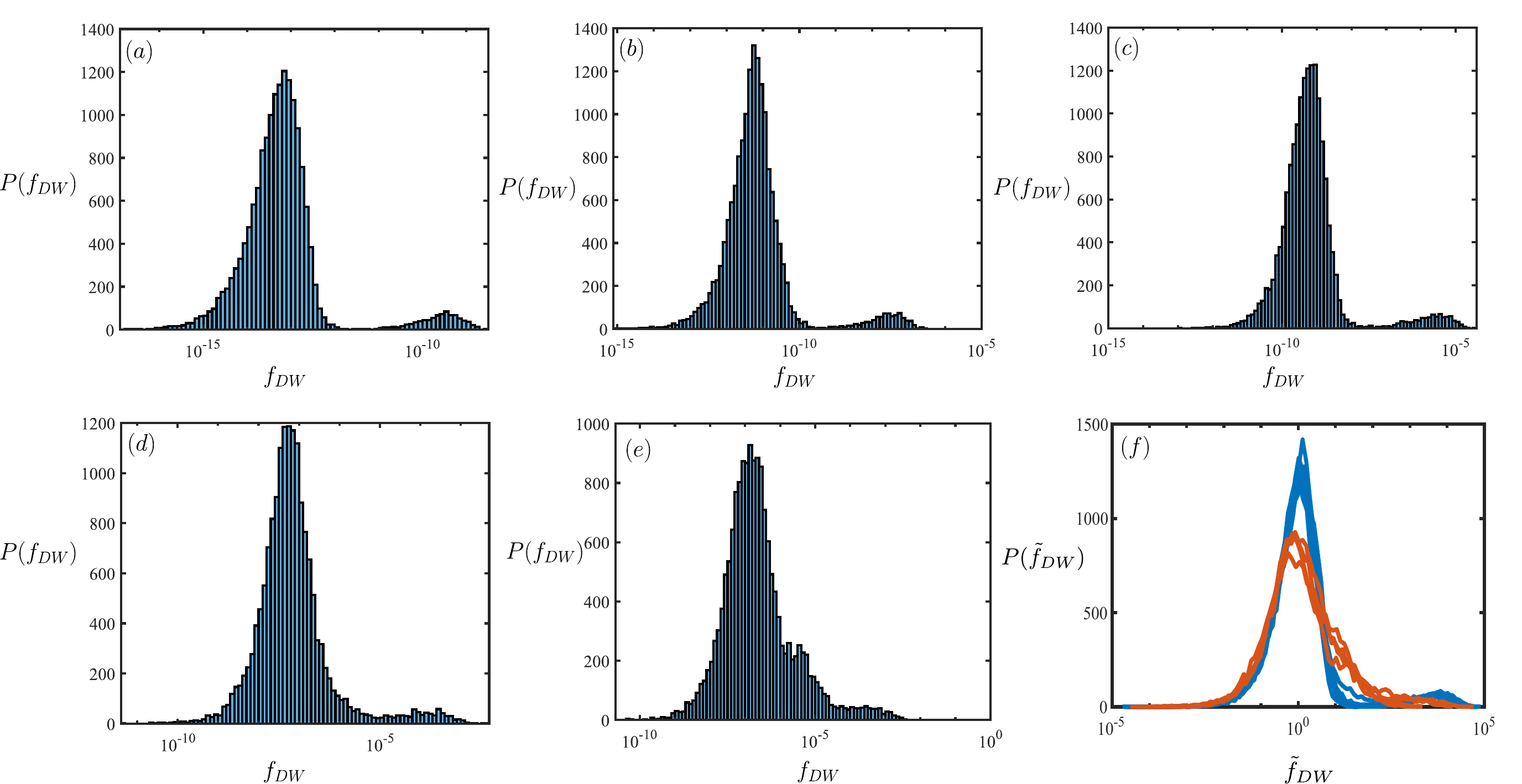}
	\caption{Distribution of Debye-Waller (DW) factor for systems initially at pressure $p=10^{-4}$ for strength of active forces both below and above the critical active force $f_{c}$. Active force magnitudes are given as follows (a) $f = 10^{-9}$, (b) $f = 10^{-8}$, (c) $f = 10^{-7}$, (d) $f = 2\times 10^{-6}$ and (e) $f = 3\times 10^{-6}$. At pressure $p=10^{-4}$, the critical active force is measured to be $f_c = 2.24 \times 10^{-6}$. (f) Distribution of rescaled and shifted DW factors $\tilde{f}_{\rm DW}$. Blue lines denote distributions with $f<f_c$ exhibiting a clearly bimodal distribution characteristic of a mechanically stable vibrating backbone accompanied by floaters while orange lines denote distributions for $f>f_c$ for which bimodality has disappeared. }
	\label{fig:dw_dist}
\end{figure}

Here, we concluded to the existence of a finite-sized active jammed region in the vicinity of the athermal jamming point. To reach this conclusion, we defined jammed states as those possessing a positive steady-state time-averaged excess coordination. This is a natural extension of the jammed states definition used in the athermal passive case. Nevertheless, in this section, we exhibit three dynamical quantities of interest which display clear signatures of this critical transition.

To characterize the dynamics of our systems, we focus first on single particle dynamics. A first correlation function to study is the mean-squared displacement (MSD); from our simulations, it is easily measured as 
\begin{equation}
\Delta^2 (t) = \frac{1}{N} \sum_{i=1}^N \langle \left| \Delta \rv_i(t)\right|^2 \rangle
\label{seq:MSD}
\end{equation}
where $\Delta \rv_i(t) = \rv_i(t) - \rv_i(0)$ and $\rv_i(t)$ represents the time-dependent position of particle $i$. Note that the brackets in Eq.\,(\ref{seq:MSD}) represent an average over long-time trajectories. The long-time limit of the MSD is commonly called the Debye-Waller (DW) factor. 

The measure of the MSD for jammed packings is notoriously plagued by the dynamics of so-called floaters. Floaters are a small fraction of particles from the original packing which are not part of the mechanically stable backbone of the static packing. Being less connected than other particles, floaters have peculiar dynamical properties. In particular, when the systems is subject to thermal or active fluctuations, they move much more than the other particles and so their dynamics, while not representative of the overall dynamics of the system, may dominate the sum in Eq. (\ref{seq:MSD}).

One way to deal with floaters is to remove them from the sum in Eq. (\ref{seq:MSD}); to do so, we need to identify the floaters which is not an easy task in a dynamic packing. In an overcompressed static packing, floaters are identified based on force balance considerations. Namely, they are defined as particles having strictly less than 3 contacts and easily removed recursively. However, this method can not be used for either unjammed states or dynamical states (as particles may come in and out of contact). Instead, our definition of floaters follows from \cite{Ikeda2013b} where floaters are defined as particles whose DW factor $f_{\rm DW}$ is 5 times larger than the median DW factor $\bar{f}_{\rm DW}$. Once identified, the $N_f$ floaters are removed from the sum in Eq. (\ref{seq:MSD}) which thus runs to $N' = N - N_f$.

The distributions of observed DW factors $P(f_{\rm DW})$ in configurations subject to active forces is very informative. As seen in Fig.\,\ref{fig:dw_dist}, static packings at pressure $p=10^{-4}$ subject to active forces below the critical active forces display a bimodal DW factor distributions [see Fig.\,\ref{fig:dw_dist}(a)-(c)]. In these bimodal distributions, the peak at higher DW factors is much smaller in amplitude and is a clear signature of the floaters. As the driving active force is increased above the critical active force measured via the excess coordination, we observe a merging of the two peaks in the distribution which becomes unimodal  [see Fig.\,\ref{fig:dw_dist}(d)-(e)]. This is clear in Fig.\,\ref{fig:dw_dist}(f) where we represent the distribution of rescaled DW factors $\tilde{f}_{\rm DW}= f_{\rm DW}/\bar{f}_{\rm DW}$, where $\bar{f}_{DW}$ denotes the median of the distribution. 

\begin{figure}[ht!]
	\centering
	\includegraphics[width=0.48\columnwidth]{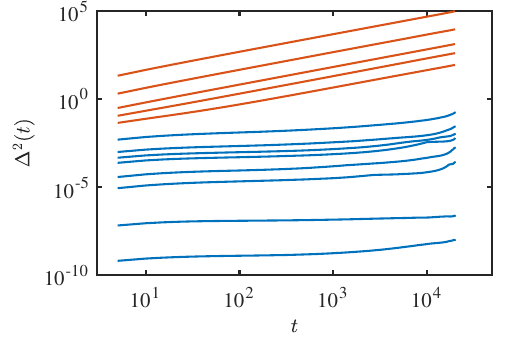}
	\includegraphics[width=0.48\columnwidth]{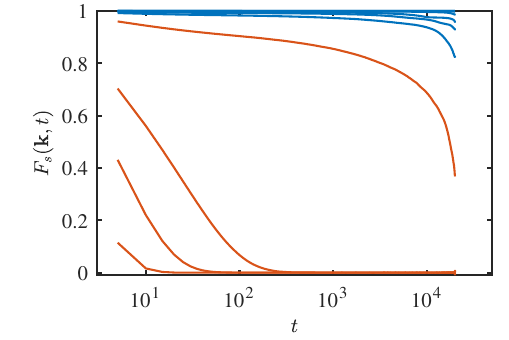}
	\caption{(a) Mean-squared displacement $\Delta^2(t)$ and (b) self intermediate scattering function $F_s(\kv,t)$ measured in systems initially at pressure $p=10^{-2}$ and subject to active forces below and above the critical active force $f_c = 0.0245$. As in Fig.\,\ref{fig:dw_dist}, color encodes the value of the active forces relative to the critical active force.}
	\label{fig:msd}
\end{figure}

Armed with a definition for the floaters, we can now compute unambiguously the mean-squared displacement $\Delta^2(t)$ for systems subjected to active forces below and above the critical active force. Fig.\,\ref{fig:msd}(a) shows that for all active forces $f<f_c$, the MSD clearly plateaus and our packings are fully caged (over the timescale of our long-time simulations). For $f>f_c$, we observe a strikingly different behavior with diffusive MSDs and we conclude that the system is unjammed passed the critical active force $f_c$. Finally, we compute the self intermediate scattering function (ISF) over the same simulations. The ISF is defined as 
\begin{equation}
F_s(\kv,t) = \frac{1}{N} \left\langle \sum_{i=1}^{N'} \exp\big[\i \kv \cdot (\rv_i(t) - \rv_i(0))\big]   \right\rangle
\end{equation}
where the wave vector $\kv$ was chosen to be proportional to the inverse of the position of the first peak in the radial pair correlation function $g(r)$. Fig.\,\ref{fig:msd} shows clearly that the ISF does not significantly decay over the full timescale of our simulations when $f<f_c$; we thus conclude to the lack of structural rearrangements over the course of these simulations. On the other hand, for $f>f_c$, we observe a very fast and sharp decay of the ISF to zero clearly denoting unjammed states. Altogether, these observations support our choice of excess coordination as a key parameter to identify the critical strength of active force for the transition.


%